\documentclass[%
 reprint,
superscriptaddress,
 amsmath,amssymb,
 aps,
]{revtex4-2}
\usepackage[T1]{fontenc}
\usepackage{graphicx}
\usepackage{dcolumn}
\usepackage{bm}
\usepackage[utf8]{inputenc}
\usepackage[version=3]{mhchem} 
\usepackage{xcolor} 
\usepackage{float}
 
 \usepackage[normalem]{ulem} 

\begin{document}

\preprint{APS/123-QED}

\title{A compact laser-plasma source for high-repetition-rate bi-modal X-ray and electron imaging}

\author{Angana Mondal}
\affiliation{Tata Institute of Fundamental Research, Homi Bhabha Road, Colaba, Mumbai 400005, India}
\author{Ratul Sabui}
\affiliation{Tata Institute of Fundamental Research,
Gopanapally, Serilingampally, Telangana 500046, India. \\
}
\author{Sonali Khanna}
\affiliation{Tata Institute of Fundamental Research,
Gopanapally, Serilingampally, Telangana 500046, India. \\
}
\author{S V Rahul}
\affiliation{Tata Institute of Fundamental Research,
Gopanapally, Serilingampally, Telangana 500046, India. \\
}
\author{Sheroy Tata}
\affiliation{Tata Institute of Fundamental Research,
Gopanapally, Serilingampally, Telangana 500046, India. \\
}
\author{M Anand}
\affiliation{Tata Institute of Fundamental Research,
Gopanapally, Serilingampally, Telangana 500046, India. \\
}
\author{Ram Gopal}
\affiliation{Tata Institute of Fundamental Research,
Gopanapally, Serilingampally, Telangana 500046, India. \\
}
\author{M Krishnamurthy}
\affiliation{Tata Institute of Fundamental Research, Homi Bhabha Road, Colaba, Mumbai 400005, India}
\affiliation{Tata Institute of Fundamental Research,
Gopanapally, Serilingampally, Telangana 500046, India. \\
}

\begin{abstract}
\noindent
Bright sources of high-energy X-rays and electrons are indispensable tools in advanced imaging. Yet, current laser-driven systems typically support only single-modality imaging, require complex infrastructure, or operate at low repetition rates.
Here, we demonstrate a compact, table-top laser-plasma source capable of generating synchronized electron and X-ray pulses at 1 kHz using just 2 mJ per pulse. A structured methanol droplet target enables quasi-single-shot electron radiographs and broadband, energy-resolved X-ray images, facilitating bi-modal imaging of both metallic and biological specimens.
We achieve resolutions of 13.6 $\mu$m for electrons and 21 $\mu$m for X-rays, and demonstrate tomographic reconstruction using 35 projections. This compact platform rivals large-scale petawatt systems in resolution and brightness, while remaining scalable and accessible for high-throughput imaging in materials science and biomedicine.

\end{abstract}

\maketitle




\section*{Introduction}
\noindent High-energy electron and X-ray imaging are foundational techniques in fields spanning biology, archaeology, materials science, and medical diagnostics.\cite{Ruska, Polman2019, Sakdinawat2010}. While hard X-rays ($>$10 keV) penetrate bulk structures to reveal internal detail \cite{Cole2015, Glinec2005, COlemouse}, electron imaging is ideal for surface morphology, contamination, and compositional mapping\cite{Zhou2014, emicro2012,jaikumar2021analysis,Nedala2024}. Combining both modalities in a single system offers powerful, complementary insight — enabling bi-modal radiography with enhanced spatial resolution and structural specificity across depth and surface. Despite the promise of dual-modality imaging, most existing systems are single-modality, with practical bi-modal implementations remaining rare due to challenges in compact source integration and beam synchronization.\\
\noindent Traditional systems rely on photocathodes for electron emission, with subsequent bremsstrahlung or characteristic X-ray production~\cite{behling2021modern}. While laser–matter interactions (LMIs) have enabled pulsed X-ray and particle sources with micrometer-scale source sizes and single-shot capability\cite{Geddes2004, Kneip2010}, both conventional and LMI-based approaches typically derive X-rays from high-energy electrons. As a result, true bi-modal radiography using separate yet concurrent electron and X-ray beams remains elusive — especially in high-repetition-rate, compact configurations.\\
\noindent Operating at just 2 mJ per pulse and 1 kHz repetition rate, our system produces synchronized dual-emission with source sizes of $\leq$13.6 $\mu$m for electrons and $\leq$ 21 $\mu$m for X-rays, with corresponding spectral ranges of 200 keV–1 MeV (electrons) and 50–210 keV (X-rays). The platform supports near-single-shot electron radiography, broadband X-ray imaging with exposure times of 10 s, and tomographic reconstruction with 35 projections. As a proof of principle, we demonstrate bi-modal imaging of a mouse paw, revealing internal skeletal structure via both modalities. The experimental layout and structured droplet geometry are illustrated in Fig. \ref{fig:one}, shown at the beginning of the \textit{Results} section to guide the reader before diving into characterization.\\
\noindent In contrast to prior LMI-based electron sources requiring ultra-relativistic intensities (10$^{18}$ W/cm$^2$)\cite{lwfa, Cole2015, Kneip2010}, our shaped-droplet platform delivers comparable or superior brightness at significantly lower intensities (2–5 × 10$^{16}$ W/cm$^2$). Conventional wakefield or solid-target schemes typically suffer from limited repetition rates, beam divergence, and large source sizes \cite{TaPhuoc2012, Powers2014, Glinec2005, Ben2011, DOPP2016}. Our source, operating stably at 1 kHz, provides distinct advantages in resolution, throughput, and compact implementation. A quantitative comparison of emission parameters, repetition rates, and operational conditions is provided in Table 1, underscoring the competitive performance of our source at significantly lower intensities.\\
\noindent The source design harnesses two-plasmon decay (TPD) dynamics in a pre-structured, 15 $\mu$m methanol droplet. To achieve efficient high-energy particle generation, we employ a dual-pulse configuration where an initial pulse (2$\times$ 10$^{15}$ W/cm$^2$) shapes the droplet, followed 4 ns later by a second pulse (4$\times$ 10$^{16}$ W/cm$^2$) that excites plasma waves via TPD \cite{TJMBoyd1986} in the droplet. This configuration enhances emission by an impressive factor of 40–50 (Fig. S1), significantly outperforming single-pulse interactions that produce negligible high-energy output. The onset of TPD leads to wave-breaking and subsequent emission of relativistic electron beams at $\pm$ 50$^\circ$ to the laser polarization \cite{Mondal2024}. The extended plasma gradient reduces the TPD threshold, enabling MeV-scale emission using sub-25 fs pulses—performance previously limited to petawatt-class lasers. We characterize this mechanism and demonstrate its utility for compact, high-resolution electron, X-ray, and bi-modal imaging.
\vspace{-11pt}
\section*{Results}
\begin{figure*} [t!]
\includegraphics[width=0.9\textwidth]{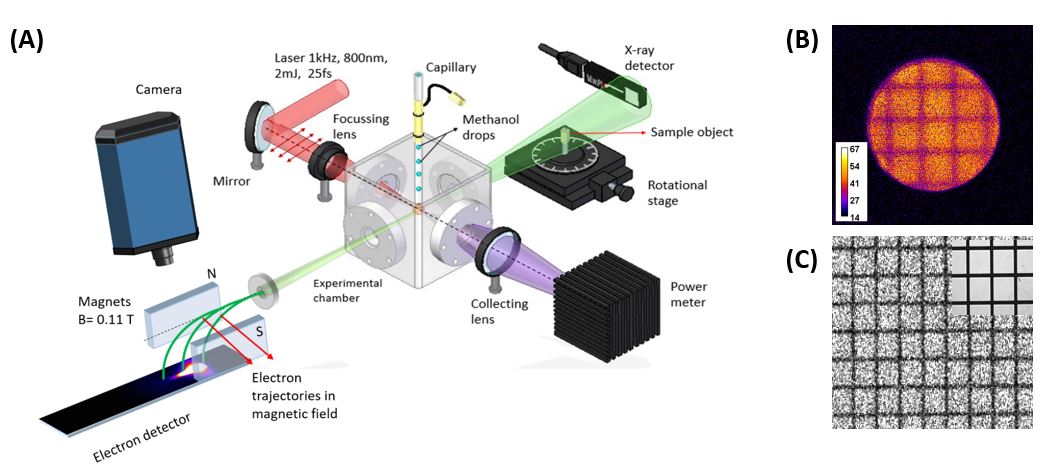}
\centering
\caption{\textbf{(A)} A schematic of the experimental set-up shows the laser focused on a 15 $\mu$m Methanol droplet with a lens of 15 cm focal length. The electron and X-ray imaging are performed at 90 $^\circ$ to the laser incidence. For X-ray imaging and tomography the samples are placed outside the experimental chamber and the images are acquired using a MINPIX detector\cite{minipix1,minipix2}. \textbf{(B)} Electron radiograph of a 41 $\mu$m Ni mesh and spacing of 321  $\mu$m obtained on LANEX for a camera exposure of 3 ms. \textbf{(C)} X-ray radiograph  of a 41 $\mu$m Ni mesh and spacing of 321  $\mu$m obtained on MINPIX detector for a camera exposure of 60 s. Inset shows an optical image of the mesh.}
\label{fig:one}
\end{figure*}
\noindent Figure \ref{fig:one} (A) presents a schematic of the compact, table-top plasma source. A stream of methanol is injected through a 10 $\mu$m glass capillary under pressure, and broken into a uniform train of 15 $\mu$m droplets via a 1 MHz piezoelectric vibrator. These droplets are irradiated by laser pulses (2 mJ, 800 nm, 25 fs) at a repetition rate of 1 kHz, producing a high-brightness electron and X-ray plasma source.
This configuration enables stable, continuous target delivery synchronized with high-repetition-rate laser operation, forming the core platform for all subsequent imaging results.\\
\noindent Without a pre-pulse, single-pulse laser-droplet interactions yield minimal high-energy electron or X-ray emission. However, introducing a collinear pre-pulse ($\sim$5$\%$ of the main pulse energy), timed 4 ns ahead, with precise spatial overlap between the two pulses maximizes both the electron and X-ray yields. The resulting emissions are captured via a LANEX screen with  charge-coupled device (CCD) camera for electrons, and a MINIPIX detector for X-rays\cite{minipix1,minipix2}.\\
\noindent Representative electron and X-ray radiographs of a 41 $\mu$m Ni mesh (321 $\mu$m pitch) are shown in Fig. \ref{fig:one}(B, C). Electron images were captured with 3 ms exposure using a LANEX screen, while X-ray images required a longer 60 s exposure using the MINIPIX detector. The X-ray inset (inset of Fig. \ref{fig:one} (C)) displays the corresponding optical image. These images demonstrate the ability of the droplet source to generate clear radiographs of micron-scale structures with high spatial resolution. 
\onecolumngrid
\begin{center}
\begin{tabular}{ | m{3.25cm} | m{3.25cm}| m{3.25cm} | m{3.25cm} | m{3.25cm} |  } 
  \hline
 Journal& Emission strength  & Energy Range & Intensity & Target\\ 
  \hline
*Uhlig,\textit{et al} \cite{uhlig} & 520 pC/s/sr/MeV & peaking at 300 keV and extending to 500keV & 10$^{15}$-10$^{18}$ W/cm$^{2}$ & Liquid Jet \\ 
\hline
Fiester, \textit{et al} \cite{Feister17} & 200 pC/s/sr/MeV & peaking a 1 MeV & 5.4 $\times$ 10$^{18}$ W/cm$^{2}$ & Liquid droplet\\ 
\hline
He, \textit{et al} \cite{He}, 
 & 12 fC/bunch (or pulse) & at 95 keV & 7.4 mJ/pulse
Focused to 2.5 $\mu$m spot size ($>$10$^{18}$ W/cm$^{2}$ )
& LWFA experiments  (500 Hz)\\ 
  \hline
Salehi, \textit{et al}\cite{Salehi17} 
 & 10 fC/bunch (or pulse)  & $>$0.5 MeV & 1.3 mJ/pulse 9 $\mu$m spot size(FWHM)
& LWFA experiments (1 kHz)\\
  \hline
He, \textit{et al}\cite{He_2013}
 & $\sim$10 fC/bunch (or pulse)  & 100 keV & 3$\times$ 10$^{18}$ W/cm$^{2}$  & LWFA experiments (0.5 kHz)\\
  \hline
Present work
 & $\sim$2200 pC/sr/s/MeV ($\sim$ 417 fC/bunch in the two electron lobes) & peaking at 200 keV & 2-5 $\times$ 10$^{16}$ W/cm$^2$(2 mJ/pulse 11 $\mu$m spot size(FWHM)) & Liquid droplet with dynamically altered shaped (1 kHz)\\
\hline
\end{tabular}
\end{center}
{\textbf{Table 1: }Comparison of electron and X-ray source properties from this work and previous laser–plasma experiments.*All cited results use 40 mJ, 10 Hz lasers; one notes similar output with 3 mJ at kHz rate.}
\pagebreak
\begin{figure*} [t!]
\centering
\includegraphics[width=\textwidth]{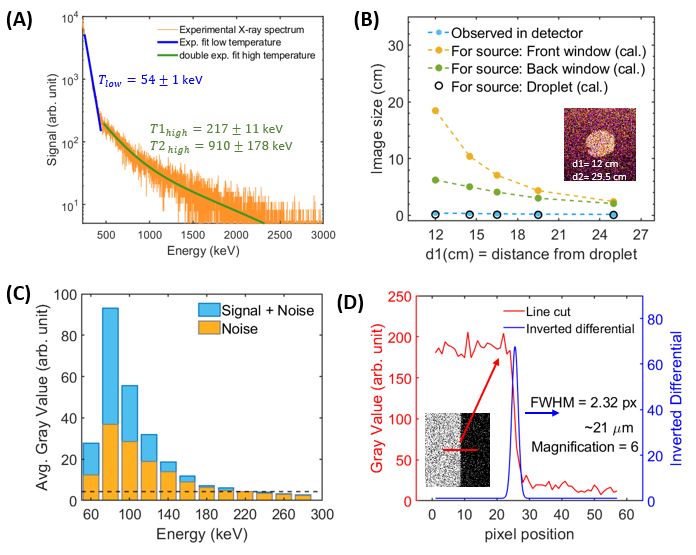}
\caption{\textbf{(A)} X-ray spectrum obtained with a NaI(Tl) detector filtered with 6 mm Pb at 4$\times$ 10$^{16}$ W/cm$^2$ for an exposure time of 900 s. The filter transmission has been accounted for in this spectrum. \textbf{(B)} Comparison of the measured variation of X-ray pin hole image diameter as a function distance between the pinhole and the droplet and the calculated image sizes assuming different source origins. The inset shows a typical image obtained on the MINIPIX detector when a Pb aperture of 1 mm diameter and 1 mm thickness is placed. d1 denotes the distance between the droplet and the pinhole, while d2 is the distance between the pinhole and the detector \textbf{(C)} Comparison of the X-ray signal within the pinhole (X-ray image spot - blue bar graph) and the background (X-ray signal beyond the spot made by the aperture- yellow bar graph) as a function of X-ray energies. The y-axis represents the total counts obtained within a 20 keV bin width, and the x-axis represents the central bin energy.\textbf{(D)} Line-cut of the image obtained from penumbral imaging of a knife-edge.}
\label{fig:two}
\end{figure*}

\twocolumngrid
\vspace{-11pt}
\section*{Characterization of the X-ray source}

\begin{figure*} [t!]
\centering
\includegraphics[width=\textwidth]{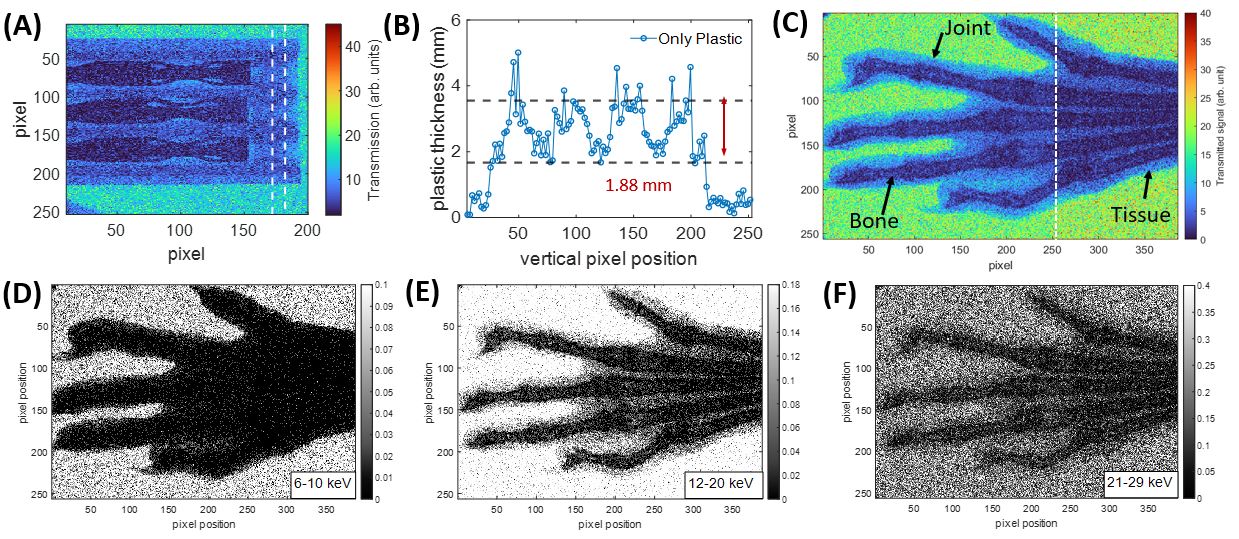}
\caption{\textbf{(A)}X-ray images of a 3-pin Molex connector encased in polyester, obtained for a duration of 100 s respectively for an energy range of 10-100 keV. \textbf{(B)} Plastic thickness (within the dashed rectangle in Fig. \ref{fig:four} (A)) calculated from the X-ray transmission with known material density of  1.38 g/cm$^3$, as a function of the vertical position for 8 keV. \textbf{(C)} X-ray radiograph of a mouse paw at a reduced laser intensity of $3 \times 10^{15}$ W/cm$^2$ summed over the energy range 5-50 keV, for an exposure time of 15 minutes. As the object exceeded the size of the detector, the image was acquired in two consecutive sections, of 15 mins acquisition, each. The white dashed line demarcates the boundary of these two images. An factor of 1.13 and 1.35 have been multiplied to the left and the right sections, respectively, for an uniform brightness over the entire image. \textbf{(D)-(F)}. Energy resolved X-ray images of panel (C), within different energy windows, demonstrate the capability of the broadband source for selective imaging of skin, tissues and bones respectively.  
}
\label{fig:three}
\end{figure*}
\noindent To characterize the X-ray source, we first identify the emission processes. As shown in Fig. \ref{fig:two} (A), two types of bremsstrahlung contribute: (i) low-energy electrons trapped within the droplet, and (ii) high-energy electrons interacting with the metallic chamber. Although both contribute to emission, the first offers smaller source size—crucial for high-resolution, lens-less imaging ~\cite{smallsource}.\\
\noindent We isolate the droplet-emitted component using a pinhole imaging setup: a 1 mm Pb sheet with a 1 mm aperture filters X-rays below 200 keV. Images acquired at varying distances from the source (d1) and pinhole-to-camera (d2) allow source localization. The experimental geometry is detailed in Fig. S2. \\
\noindent The detected image size is described by Equation (1), where demagnification (d1/d2) and aperture blur (w(d1 + d2)/d1 for an aperture size w are factored.  Figure \ref{fig:two} (B) shows that calculated spot sizes for the droplet match closely with measured images, confirming that the emission arises from the droplet, not secondary chamber interactions. In contrast, X-rays from chamber windows create diffuse backgrounds. Energy-resolved analysis (Fig. \ref{fig:two}(C)) confirms that X-rays up to $\sim$210 keV originate from the droplet; beyond this, the emission becomes uniform. However, since MINIPIX efficiency drops to 2$\%$ beyond 50 keV, this cut-off may be detector-limited.\\

\begin{equation}
\sigma_{image} = ((\sigma_{source} /demag)^2+ \sigma_{blur}^2)^{1/2}
\end{equation}

\noindent
 We further estimate source size using penumbral imaging~\cite{Armstrong2019},  as shown in Fig. \ref{fig:two} (D). The transition gradient across a 100 $\mu$m thickness and about 321 mm radius, yields a source size of $\sim$ 21 $\mu$m, —comparable to the droplet diameter and notably smaller than 32 $\mu$m resolution reported from betatron X-rays at 6 $\times$ 10$^{18}$ W/cm$^2$~\cite{COlemouse}. In addition, due to the available magnification constraint, the calculated resolution of 21 $\mu$m is only an upper limit.
 \begin{figure*} [t!]
\centering
\includegraphics[width=\textwidth]{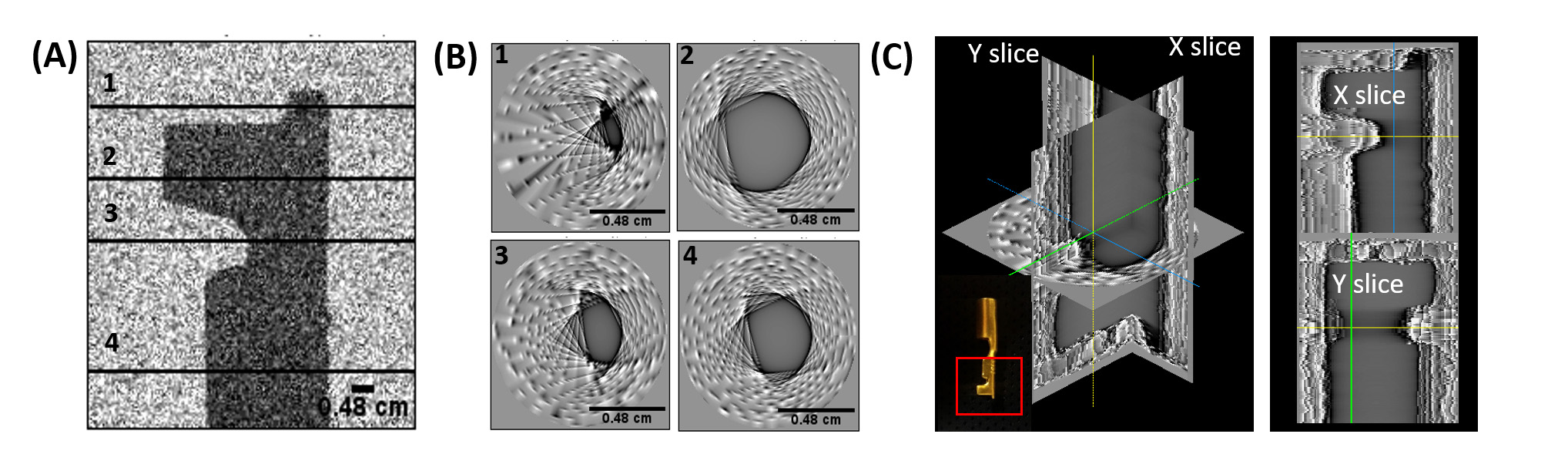}
\caption{\textbf{(A)} X-ray projection image of a metal object  obtained for 100 s acquisition time.  Four cross-sectional slices from reconstructed tomogram taken at different heights are shown on the right\textbf{(B)}. \textbf{(C)}The X-slice and Y-slice projection of the tomogram reconstructed from 35 projections spanning from 0$^\circ$ to 350$^\circ$. The blurry features in the reconstruction appear owing to less number of projections. The rectangle on the actual object shown in the inset denotes the area imaged with X-rays.
}
\label{fig:four}
\end{figure*}
\vspace{-11pt}
\section*{Energy-Resolved Radiography with Broadband X-rays}
\noindent The broadband nature of the droplet-based X-ray source, though initially perceived as a limitation, enables versatile imaging when combined with energy-resolved detectors such as MINIPIX. As shown in Fig. \ref{fig:three} (A), a  100 s exposure at 4$\times$ 10$^{16}$ W/cm$^2$, yields a clear radiograph of a 3-pin Molex connector encased in polyester. Despite the polymer housing, internal structural variation is evident from opacity gradients. In the dashed region, these periodic contrast patterns correspond to vertical thickness thickness variations of 2 $\pm$ 0.1 mm of the polyester casing.\\
\noindent Using known attenuation coefficients of polyester at 8, 10 and 12 keV\cite{Xrayplastic}, we estimate a vertical thickness variation of $\sim$ 2.5 $\pm$ 0.5 mm (Fig. \ref{fig:three} (B)). The close agreement between physical and radiographic measurements confirms the system’s capability to resolve millimeter-scale features in low-Z materials, suggesting its suitability for high-contrast imaging in polymer-encased electronics and biomedical soft tissues.\\
\noindent To demonstrate biological relevance, a preserved mouse paw was imaged using broadband X-rays (6–50 keV) over a 15-minute exposure. The large exposure is chosen as the image is acquired at a reduced intensity of 3 $\times$ 10$^{15}$ W/cm$^2$ with a stretched pulse of 300 fs,  without any contrast enhancement assays. Figure \ref{fig:three} (C) reveals both soft tissue and skeletal structure without the use of contrast agents. Skin layers, joints, and digit bones are all discernible, confirming that the droplet source enables high-resolution biological imaging, even in unprepared specimens.\\
\noindent Leveraging MINIPIX’s energy-resolved imaging mode, we further acquired segmented radiographs within narrow energy bands. Figure \ref{fig:three} (D-F) show the same mouse paw imaged at 6–10 keV, 12–20 keV, and 20–29 keV respectively. At low energies, soft tissue dominates; at intermediate energies, both skin and bone are visible; and at higher energies, tissue becomes nearly transparent, revealing only skeletal features. These transitions confirm spectral tunability for imaging contrast, without complex switching mechanisms, mimicking the principle of dual-energy CT in a much simpler form.\\
\noindent A energy-resolved video (Supplementary Movie 1) visualizes the energy-dependent transparency progression across the sample, highlighting the natural evolution of contrast with increasing photon energy.\\
\noindent As a secondary emission process, X-ray yield is inherently lower than direct electron emission. Nonetheless, assuming isotropic emission, we achieve a photon flux of $\sim$ 9.1 $\times$ 10$^6$ photons/sr s in the 50–210 keV range (post-filtering) at 25 fs pulse width and intensity of 4$\times$10$^{16}$ W/cm$^2$, the measured photon flux in the range of 50-210 keV.  As shown in Fig. S3, this allows exposures as short as 10 s, depending on sample size and attenuation.\\
\noindent The system’s resolution, currently 21 $\mu$m, is derived from penumbral analysis (Fig. \ref{fig:two}(D)). Higher resolution may be achieved with optimized detector-pixel size and spot collimation. Further, higher laser repetition rates (10–100 kHz) could reduce acquisition time significantly and allow live-frame applications.\\
\noindent Finally, we demonstrate proof-of-principle 3D tomography of a metallic micro-connector using the same X-ray source. Acquiring 35 angular projections across 0$^\circ$ to 350$^\circ$°, with 100 s exposure per angle (10–100 keV range), Reconstruction using Octopus Imaging Software~\cite{Dierick_2004} yields transverse and longitudinal slices of the object, shown in Fig. \ref{fig:four}(A–C).\\
\noindent Despite sparse angular sampling, the sample geometry is captured with a spatial error of $\sim$60 $\mu$m, limited primarily by stage precision and beam divergence. This establishes the feasibility of 3D imaging, marking a rare demonstration of hard X-ray tomography from a droplet-based laser–plasma source operating well below petawatt intensities.
\vspace{-11pt}
\section*{Characterization and imaging of the electron source}

\begin{figure*} [t!]
\centering
\includegraphics[width=0.8\textwidth]{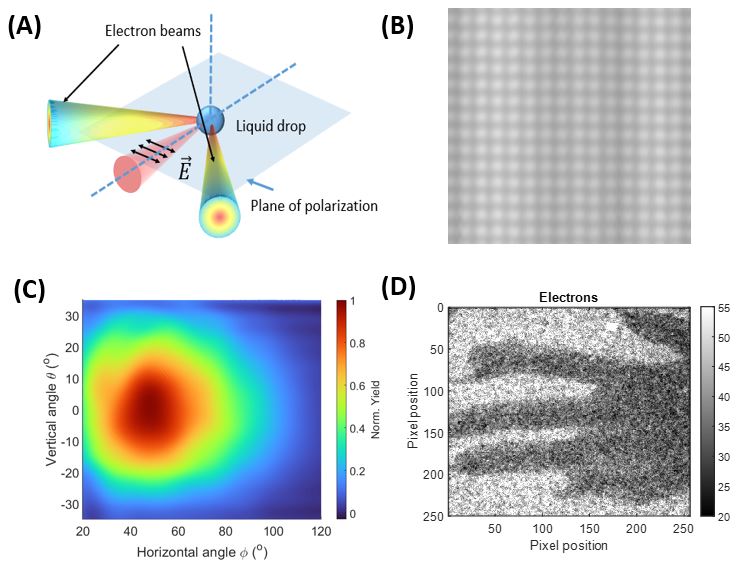}
\caption{\textbf{(A)} Schematic of the electron angular distribution, as emitted from the droplet target, at 4$\times$10$^{16}$ W/cm$^2$.  The electrons are emitted as twin beams in the plane of the laser polarization, directed at an angle of $\sim \pm 50 ^\circ$ with respect to the backward laser propagation direction. \textbf{(B)} An electron transmission image of the $\sim$13.6 $\mu$m Ni mesh acquired at 4$\times$10$^{16}$ W/cm$^2$ on an image plate for an exposure of 30 s. Further detail of the electron emission direction is presented in Ref \cite{Mondal2024}. \textbf{(C)} Electron angular distribution is captured by placing image plates around the droplet source, on a circle of 3.2 cm radius. The laser propagates into the plane at (0$^\circ$,0$^\circ$) ($\theta$,$\phi$) position, with $\theta= 0^\circ$ marking the plane of polarization, where $\phi$ is the angle in the horizontal plane and $\theta$ is the angle in the vertical plane. \textbf{(D)} Electron radiograph of the mouse paw (left panel in Figure \ref{fig:four} (C)) at a reduced laser intensity of $3 \times 10^{15}$ W/cm$^2$, acquired simultaneously with the X-rays, for an exposure time of 15 minutes.}
\label{fig:five}
\end{figure*}

\noindent In addition to broadband X-rays, the same laser–droplet interaction produces energetic electron beams, offering a complementary modality for surface-sensitive, high-resolution imaging. The droplet source generates a broadband electron spectrum extending beyond 4 MeV, with the dominant component between 200 keV and 1 MeV (Fig. S4). Electrons are emitted in a twin-beam geometry at $\pm$ 50 $^{\circ}$ relative to the laser axis, (Fig. \ref{fig:five} (A)), a characteristic signature of the two-plasmon decay (TPD) mechanism \cite{Mondal2024}.\\
\noindent From image plate measurements and angular distribution (Fig. \ref{fig:five} (C)), the total yield is estimated to be $\sim$440 pC/sr/s or $\sim$ 3 $\times$ 10$^6$ electrons/sr/shot, within the energy range of 200 keV to 6 MeV. The energy conversion efficiency from the above measurement, excluding the loss via bremsstrahlung, is calculated to be 4 $\times$ 10$^{-3}$ $\%$, which is $\sim$ 10 $\times$ higher yield and $\sim$ 100 $\times$ higher conversion efficiency than prior droplet studies at 10$^{16}$ W/cm$^2$~\cite{dropcompare}. This performance highlights the effectiveness of our low-intensity, high-repetition-rate configuration, enabling electron yields comparable to laser wakefield setups but at far milder operating conditions.

\noindent To measure the electron source size, we implement the simplest technique of determining minimum achievable imaging resolution. The effective source size is characterized by direct transmission imaging of a Ni mesh ($\sim$ 13.6 $\mu$m and grid spacing of $\sim$ 40 $\mu$m, SEM image in Fig. S5) using an image plate (Fig. \ref{fig:five} (B)). The mesh structure is sharply resolved with a 30 s exposure and a 22.5$\times$ magnification factor, implying an upper limit of 13.6 $\mu$m for the source size—on par with conventional thermionic sources\cite{SEM2003}. Importantly, even at 30,000-shot integration, the source jitter remains below the droplet size, confirming high temporal stability. Compared to existing laser-driven electron sources ~\cite{erad}, , this approach yields $\sim$5$\times$ better resolution under longer exposures, underscoring the robustness and consistency of the source.\\
\noindent While temporal characterization has not been performed experimentally, the small source size implies minimal droplet expansion, and hence ultrashort emission duration. Particle-in-cell (PIC) simulations under experimental conditions estimate pulse widths of $\sim$43 fs for 200–300 keV electrons, shortening to a few femtoseconds for higher energies (Fig. S6(B)- (D)).\\
\noindent To demonstrate quasi-single-shot imaging, a LANEX screen captured electron radiographs of a 41 $\mu$m  Ni mesh at 3 ms exposure (Fig. \ref{fig:one}(B)). The unsynchronized camera trigger implies an effective exposure of at most two laser shots. The mesh is clearly resolved, validating the system's ability to generate images with only one or two pulses. \\
\noindent A 200 G magnet placed near the detector deflects the image entirely (Fig. S7), confirming the electrons as the source of the radiograph. The usable energy window is defined by the LANEX detection threshold ($\sim$ 200 keV  ~\cite{LANEX}) and the observed cut-off at $\sim$1 MeV (Fig. S4). Beyond this energy range upto 1 MeV, the electron yield falls below the detection limit of single-shot acquisition.\\
\noindent Notably, current imaging is performed within a 20$^\circ$ angular collection cone centered at 90$^\circ$ to the laser, capturing only one-third of the peak yield. Aligning the detector at the 50$^\circ$ emission angle would reduce exposure times to 1 ms or less, enabling true single-shot performance.
\vspace{-11pt}
\section*{Bi-modal imaging capability with X-rays and electrons}
\noindent A key advantage of the laser–droplet interaction is its ability to simultaneously generate both X-rays and electrons from the same physical site and trigger. This inherent synchronization allows imaging of the same sample with both modalities—without requiring realignment or repositioning.\\
\noindent The MINIPIX detector responds to both photons and charged particles, including electrons, protons, and alpha particles \cite{Vavrik2014, LOpalka2012}. Incident X-rays typically register as single- or double-pixel hits, whereas electrons produce elongated, curly tracks across multiple pixels, as shown in Fig. S8.\\
By separating detection events based on track length, we extract two distinct radiographs from a single acquisition: the X-ray radiograph (Fig. \ref{fig:four} (C)) and the corresponding electron radiograph (Fig. \ref{fig:five} (D)). While both images show the internal structure of a mouse paw, notable differences in contrast and sharpness are observed.\\
\noindent In the electron image (Fig. \ref{fig:five} (D)), the contrast between bone and soft tissue is diminished compared to the X-ray image, primarily due to increased electron scattering in bulk tissue. Additionally, this image was acquired at a lower laser intensity (3$\times$10$^{15}$ W/cm$^2$), significantly reducing the high-energy ($\geq$ 500 keV) electron population and thus lowering penetration depth. As a result, the electron contrast—governed by the preferential transmission of high-energy electrons—is weaker.\\
Moreover, the electron radiograph is not energy-resolved and represents a cumulative projection of all electrons registering track lengths $\geq$ 5 pixels. With further development of calibration models for track length vs energy, such dual-mode datasets could enable co-registered, depth-aware reconstructions and energy-dependent contrast analysis, potentially enhancing internal feature discrimination.\\
\noindent These results establish the feasibility of synchronized, energy-selective bi-modal imaging using a compact laser–droplet system—offering a practical path toward integrated diagnostics at high repetition rates.
\vspace{-11pt}
\section*{Discussion and Conclusion}
\noindent In this work, we demonstrated a compact, high-repetition-rate laser–plasma platform capable of generating synchronized, broadband electron and X-ray pulses from a shaped methanol droplet target. The X-ray spectrum spans from a few keV up to 210 keV, while the electron energy peaks between 200 keV and 1 MeV. Without the need for petawatt-class lasers, we achieve spatial resolutions of 21 $\mu$m for X-rays and 13.6 $\mu$m for electrons—rivaling or exceeding existing sources in brightness and image fidelity.\\
\noindent Our system supports fast-acquisition X-ray imaging and near-single-shot electron radiography using only 2 mJ pulses at 1 kHz. We further demonstrate energy-resolved X-ray segmentation, soft-tissue imaging, and 3D tomographic reconstruction using just 35 angular projections. Bi-modal imaging of biological specimens is validated via a synchronized acquisition of both modalities, showing internal skeletal structure using X-rays and surface sensitivity with electrons.\\
\noindent Compared to existing laser–plasma platforms, our system operates at 10–100× lower intensities while delivering competitive output. A comparative summary of performance metrics is provided in Table 1. These results highlight the potential of our source for compact, high-throughput imaging across multiple domains. 
\vspace{-11pt}
\section*{Outlook}
\noindent Looking ahead, the integration of synchronized X-ray and electron sources within a single platform opens the door to comprehensive radiographic diagnostics. X-rays provide structural insights at depth, while electrons enable surface-sensitive imaging. Calibrating electron track length to energy could unlock energy-resolved imaging akin to dual-energy CT, enhancing material contrast without the need for contrast agents.\\
\noindent Beyond imaging, the droplet-based source holds promise for element-specific analysis via electron-induced X-ray fluorescence, as well as time-resolved studies of charge and density evolution in laser–plasma interactions. With MHz droplet generation already demonstrated, scaling the system to 10–100 kHz laser operation is feasible using current technology. These capabilities position the droplet source as a powerful, cost-effective tool for medical diagnostics, semiconductor inspection, and in situ materials analysis. This work lays the foundation for accessible, lab-scale multi-modal imaging platforms operating at high repetition rates. 
\vspace{-11pt}
\section*{Methods}
\subsection*{Pre-pulse conditions for source generation}
\noindent Enhancement of the electron and X-ray generation from the liquid droplet is observed in the presence of a pre-pulse having 5\% of the main-pulse energy (2 mJ), incident 4 ns preceeding the main pulse. This pre-pulse, focused to a spot size of  11 $\mu$m produces a peak intensity of  2 $\times$ 10$^{15}$ W/cm$^2$, capable of generating a long-length scale plasma at the target within 4 ns. The main pulse interacts with the pre-pulse affected droplet target enhancing the yield of hard X-rays and relativistic electrons\cite{ anand2, Mondal2024}. 
\vspace{-11pt}
 \subsection*{Electron and X-ray spectrum measurement}    
\noindent The measurement of the electron energy spectrum were using an electron spectrometer (ESM) coupled to a LANEX detector. Electrons generated from the laser-droplet interaction were collimated through a 2 mm aperture, prior to the ESM. A permanent magnet of 0.11 T was used to bend these electrons onto a LANEX detector and the LANEX scintillation was captured using an 8 bit CCD detector. A band pass filter of 540 nm was used to block optical emission. The signal strength was measured as a function of the incident position on the detector an converted to the respective energies \cite{Mondal2024}. The presented electron spectra was corrected using LANEX sensitivity values, obtained in literature\cite{LANEX}. The electron spectrum is measured, at 90 $^\circ$ with respect to the incident laser beam, using a standard electron spectrometer (ESM) coupled with a LANEX or image plate(IP)\cite{LANEX, IP1, IP2}. \\
\noindent The X-ray spectrum were detected using a NaI(Tl) detector, for energy range of 100 keV-6 MeV, and apre-calibrated  MINIPIX detector for energies upto 300 keV. The NaI(Tl) detector, coupled with a multi-channel analyser was gated with respect to the main pulse, to ensure detection of X-rays only during the main pulse interaction. A 6 mm Pb filter was used to avoid pile-up in the X-ray measurements. The calibration of the MINIPIX detector was verified using the L $\alpha$ and L $\beta$ lines of Pb having energies of $\sim$ 10 keV and $\sim$ 12 keV, respectively. The NaI(Tl) detector, was calibrated using  137 Cs, 22 Na and 133 Ba radioactive gamma sources. The corrections for filter transmission has been accounted for in the reported spectrum. 
\vspace{-11pt}
\subsection*{Electron angular distribution measurement}   
 \noindent The electron angular distribution was measured by placing image plate detectors spanning from 0$^{o}$ to 360$^{o}$ in the horizontal plane and from -40$^{o}$ to 40 $^o$ in the vertical plane, at a radial distance of 3.2 cm from the droplet target. The image plates are wrapped with 110 $\mu$m Al foil to block detection of both optical and low energy electrons. Circular opening are made to ensure unobstructed entry and exit of the laser beam. As the electron emission is identical on both arms\cite{Mondal2024} and  occurs predominantly within 20$^o$ to 120$^o$ with respect to the incident beam, only one beam profile is presented in the current manuscript. For further details on exact geometry please refer to the supplement of Ref\cite{Mondal2024}.
 \vspace{-11pt}
\subsection*{X-ray and electron imaging} 
\noindent The electron and X-ray imaging are also performed at 90$^\circ$ with respect to the incident laser beam. \cite{filtertrans}. The X-ray imaging was acquired using the MINIPIX detector. A 5.5 $\mu$m Al filter was used to block optical light.The energy resolved X-ray image was obtained using the Pixetpro software, provided by the manufacturer. 
For electron imaging the sample are placed inside the target chamber and images are acquired using either a LANEX detector coupled to an 8 bit CCD camera or an image plate detector. For, bi-modal imaging of the mouse-paw, the electron tracks greater than 5 pixels was separated from the raw images of 1 s exposure, each using a MATLAB code. Accumulation of 900 such images are presented in Fig. \ref{fig:five}(D).
\vspace{-11pt}
\section*{Data Availability}
\noindent The datasets generated during and/or analysed during the current study are available from the corresponding author on reasonable request.\\

\section*{Code Availability}
\noindent The MATLAB codes used to analyse the datasets are are available from the corresponding author on reasonable request.
\vspace{-11pt}
\section*{Ethics Declaration}
All animals used for experiments were approved by the Institutional Animal Ethics Committee of Tata Institute of Fundamental Research with the protocol number 'TIFRH/2024/19'(X-ray bio-imaging of Tissues). The approval complies with The Breeding of and Experiments on Animals(Control and Supervision) Rules(1998) which follows the Prevention of Cruelty to Animals Act(1960)[CCSEA, Govt. of India].   

\section*{Author Contributions}
\noindent M.K. and A.M. conceived the idea of the laser-droplet experiments in consultation with R.G. and M.A.; A.M. conducted the droplet experiments with support from  R.S., S.K., S.V.R. and S.T. The 2D- PIC simulations done by R.S.; The manuscript was written by A.M. and M.K. with the help of the other authors.

\section*{Competing interests}
The authors declare no competing interests.

\bibliography{sample}

\end{document}


\preprint{APS/123-QED}

\title{A compact laser-plasma source for high-repetition-rate bi-modal X-ray and electron imaging}

\author{Angana Mondal}
\affiliation{Tata Institute of Fundamental Research, Homi Bhabha Road, Colaba, Mumbai 400005, India}
\author{Ratul Sabui}
\affiliation{Tata Institute of Fundamental Research,
Gopanapally, Serilingampally, Telangana 500046, India. \\
}
\author{Sonali Khanna}
\affiliation{Tata Institute of Fundamental Research,
Gopanapally, Serilingampally, Telangana 500046, India. \\
}
\author{S V Rahul}
\affiliation{Tata Institute of Fundamental Research,
Gopanapally, Serilingampally, Telangana 500046, India. \\
}
\author{Sheroy Tata}
\affiliation{Tata Institute of Fundamental Research,
Gopanapally, Serilingampally, Telangana 500046, India. \\
}
\author{M Anand}
\affiliation{Tata Institute of Fundamental Research,
Gopanapally, Serilingampally, Telangana 500046, India. \\
}
\author{Ram Gopal}
\affiliation{Tata Institute of Fundamental Research,
Gopanapally, Serilingampally, Telangana 500046, India. \\
}
\author{M Krishnamurthy}
\affiliation{Tata Institute of Fundamental Research, Homi Bhabha Road, Colaba, Mumbai 400005, India}
\affiliation{Tata Institute of Fundamental Research,
Gopanapally, Serilingampally, Telangana 500046, India. \\
}

\begin{abstract}

\end{abstract}

\maketitle
\clearpage

\twocolumngrid

\section{Supplementary Note S1 :Discussion on effect of pre-plasma}

The X-ray emission as a function of the pre-pulse energy was studied by systematically varying only the pre-pulse energy while keeping the main pulse energy constant. The X-ray emission from the droplet was enhanced by 43 times in the presence of a pre-pulse having 5-6\% of the main pulse energy, as opposed to the case of no-prepulse. Increasing the pre-pulse energy further lead to a saturation in the X-ray yield emission\cite{Mondal2024}. Figure S1 shows the comparison of the X-ray yield measured at a main pulse intensity of 1.4 $\times$ 10$^{16}$ W/cm$^2$ in the presence and absence of a pre-pulse. The spectrum were obtained using the MINIPIX detector for exposure of 140 s.

\setcounter{figure}{0} 
\begin{figure}[htbp!]
\renewcommand{\thefigure}{S\arabic{figure}}
\includegraphics[width=0.9\columnwidth]{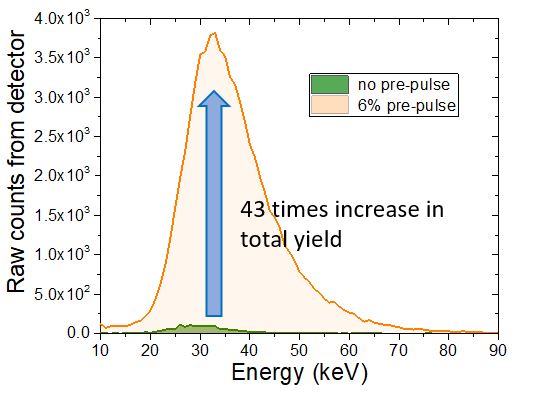}
\caption{Raw X-ray spectrum showing enhancement of emission in presence of a pre-pulse having 6\% of the main pulse energy. The main pulse intensity is about 1.4 $\times$ 10$^{16}$ W/cm$^2$. The X-ray spectrum measurements were acquired on MINIPIX detector for an exposure of 140 s.}
\end{figure}

\section{Supplementary Note 2: Pin-hole imaging geometry to determine the origin of emitted X-rays}

The pin-hole imaging geometry using 1 mm Pb pin-hole, was used to determine whether the emitted X-rays originate from he drop of the experimental chamber. Fig. S2(A) shows the pin-hole imaging scheme for source smaller greater than the aperture diameter(which essentially denotes the source being the drop). In this geometry the image dimension would be limited by the magnification of the pin-hole itself and would be smaller than the MINIPIX detector area of 1.4$\times$1.4 cm$^2$. In case the X-rays where generated from windows in the beam path(located $\pm$ 3.5 cm from the target center), the pin-hole would be smaller than the source size and the image formed would be a true magnified image of the source as shown in Fig. S2(B). The calculated dimensions of these images as shown in Fig. 2(B) would therefore exceed the detector dimension forming a constant background as opposed to a clear aperture on the MINIPIX detector.

\begin{figure}[!h]
\centering
\renewcommand{\thefigure}{S\arabic{figure}}
\includegraphics[width=0.95\columnwidth]{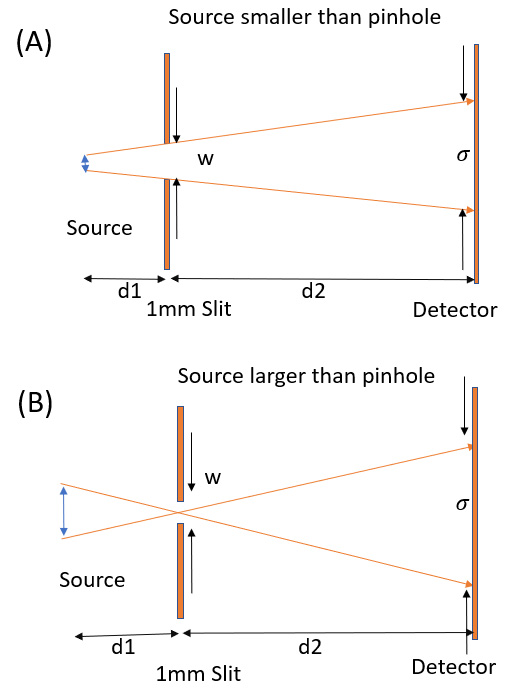}
\caption{\textbf{(A)} Schematic of pinhole imaging scheme for source size smaller than the aperture dimension. \textbf{(B)} Schematic of pinhole imaging scheme for source size larger than the aperture dimension.}
\end{figure}

\section{Supplementary Note S3: X-ray radiography of the capillary holder for 10 s exposure}
Figure S3 shows the X-ray image of the capillary connector, measured for an acquisition time of 10 s. In addition to the clearly discernible casing and metallic wired within the connector, the gradation arising from the plastic thickness variation is also visible in this short acquisition.


\begin{figure}[!h]
\renewcommand{\thefigure}{S\arabic{figure}}
\includegraphics[width=0.8\columnwidth]{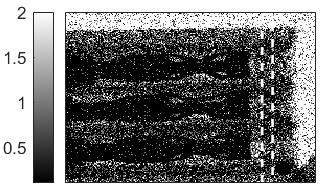}
\caption{X-ray radiograph of capillary connector acquired at $4 \times 10^{16}$ W/cm$^2$ for an exposure of 10 s. The white dashed rectangle shows the gradation of thickness in the plastic region.}
\end{figure}

\section{Supplementary Note S4: Dominant component of electron spectrum}
Figure S4 shows the dominant energy component of the electron spectrum captured on the LANEX for an exposure of 4 s. This energy component ranging from 200 keV-1 MeV is used for electron radiography. Electrons with energies beyond 1 MeV are comparatively weaker and can only be observed for larger acquisition times of several minutes \cite{Mondal2024}.

\begin{figure}[!h]
\renewcommand{\thefigure}{S\arabic{figure}}
\includegraphics[width=0.9\columnwidth]{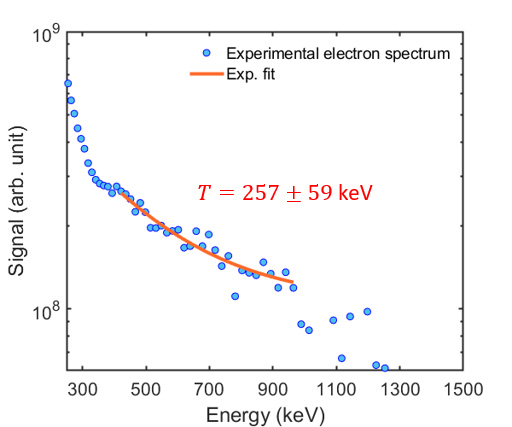}
\caption{Electron spectrum measured by the ESM coupled to a LANEX for an acquisition of 4s . The sensitivity corrected spectrum shows the dominant electron component extending up to 1 MeV, capable of  fast imaging.}
\end{figure}

\section{Supplementary Note S5: SEM image of Ni mesh}
The SEM image used for the determination of the Ni mesh thickness used for the electron radiography shown in Fig. 4(B) is shown below.

\begin{figure}[!h]
\renewcommand{\thefigure}{S\arabic{figure}}
\includegraphics[width=0.8\columnwidth]{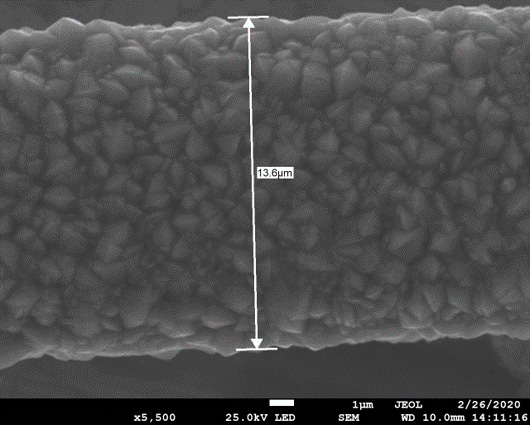}
\caption{SEM image showing the thickness of the 13.6 $\mu$m wire grid used for imaging in Fig. 4(B).}
\end{figure}

\section{Supplementary Note S6: Discussion of the 2D PIC simulations}

To obtain the temporal characteristics of the droplet generated electrons we perform 2D PIC simulations incorporating the pre-pulse modulated structure observed through shadowgraphy experiments and the expected plasma density gradient\cite{Mondal2024} .The EPOCH\cite{EPOCH} simulations are performed on a simulation box of 80 $\times$ 80 $\mu$m$^2$ divided in 3000 $\times$ 3000 cells. The droplet target is placed at the centre of the simulation box, having a density profile given by ne = 10 nc, for r$<$ 10 $\mu$m and rc $>$ 13$\mu$m, and ne/nc = 10 cos($\pi$(r-10)/10), where rc is the radius of the cup-structure, given by rc =$\sqrt{(x - (-10))^2 + y^2}$, measured from the center of the cup x = -10$\mu$m and y=0. The radius of the spherical part is given by r =$\sqrt{(x - (10))^2 + y^2}$, with its centre at x = 10$\mu$m and y=0.. The electron density in the region x$>$30 $\mu$m and rc $<$ 13 $\mu$m is given by ne/nc = 10 exp(-(13-rc)$^2$/4)). The target geometry is shown in Figure S6(A).\\
The laser  pulse is simulated by a normalized vector potential a = a0 sin($\pi$t/$\tau$ )$^2$, incident from the left boundary.  a0 corresponds to a normalized intensity $\approx$ 0.855$\sqrt{I[10^{18} W/cm^2] \lambda^2[\mu m]}$. For a peak intensity of 8.56 $\times$ 10$^{16}$ W/cm$^2$, a0=0.2. The laser pulsewidth is given by $\tau$ = 25T$_0$ = 25$\lambda_0/c =$  24.3 fs (FWHM in intensity profile). The laser has a Gaussian transverse profile and a waist radius of w$_0$ = 9.6 $\mu$m. The region surrounding the droplet is simulated by neutral atoms of Ar(having similar ionization energy to N) having a  gas density of 10$^{17}$/cm$^3$. To analyse the temporal profile of the emitted electron bunches, a virtual probe is placed at $\sim$30 $\mu$m from the droplet centre. This measures the pulse duration of the electrons emitted from the top lobe of the 50$^o$ emission. It is observed that electrons within 200-300 keV to have a bunch length of 43 fs. For higher energy electrons between 500-600 keV and 1000-1500 keV, the bunch length in seen to be reduced to 28 fs and 5 fs respectively. The electron bunch length as a function of electron energy is shown in Fig. S6(B-F), which shows that the electrons are indeed emitted in ultrashort burst of few tens of fs pulse duration. For further detains on simulations and their results, please refer to Ref \cite{Mondal2024}.\\

\onecolumngrid
\begin{figure*} [!t]
\centering
\renewcommand{\thefigure}{S\arabic{figure}}
\includegraphics[width=0.8\textwidth]{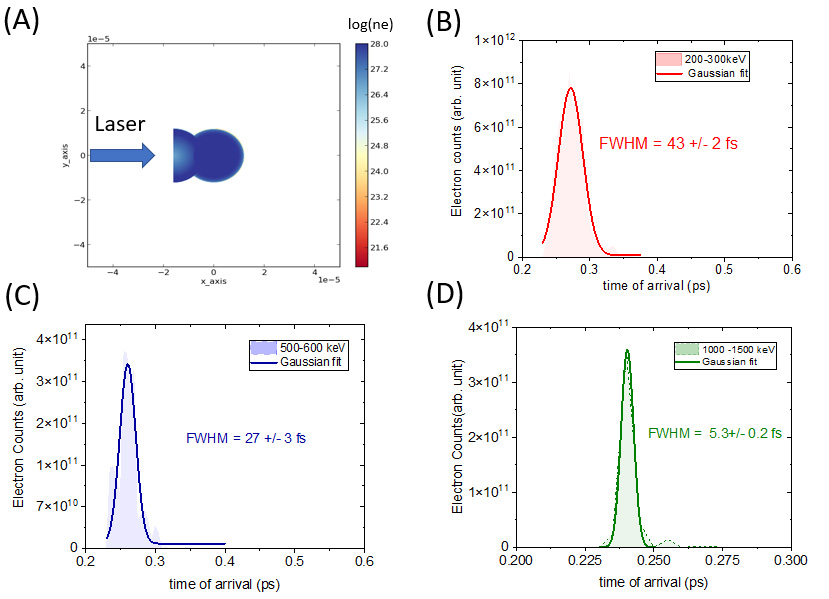}
\caption{(A) Droplet geometry considered for 2D PIC simulations using EPOCH. The x and y axis are in meters. ne is the electron density. Laser in incident from the left side of the simulation box marked by the blue arrow with the electric field polarization along the y axis.(B)-(C) The electron pulse durations for different energy electrons obtained from 2D PIC.}
\end{figure*}
\twocolumngrid

\section{Supplementary Note S7: Verification of electron imaging with LANEX}

To verify that the Ni mesh image formed on the LANEX detector (Figure 1(B)), originates from the electron, we perform back-to-back  imaging of the LANEX, in the absence (Figure S7 (A)) and presence (Figure S7(B)) of a 200 G permanent magnet. As observed in present of the magnet the Ni mesh image disappears verifying electrons to be the imaging source.
\begin{figure}[!h]
\renewcommand{\thefigure}{S\arabic{figure}}
\includegraphics[width=0.9\columnwidth]{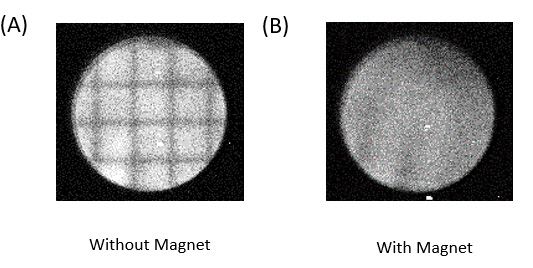}
\caption{\textbf(A) Electron radiograph of 41 $\mu$m Ni mesh, acquired for an exposure of 100 ms. \textbf(B) Image of the LANEX, under identical geometry and exposure, in presence of a 200 G magnet, near the LANEX detector.}
\end{figure}

\section{Supplementary Note S8: Separation of gamma and electron tracks using MINIPIX}
The MINIPIX detector is ideal for bi-modal imaging as it detects both X-rays and charged particles, generated by the droplet source. The separation of the captured image into X-rays and electrons can be performed post-collection based on the shape of the tracks recorded in the MINIPIX detector \cite{LOpalka2012,Vavrik2014}. For example Figure S8 (A) shows a single raw frame of Figure 2(C) of 1s exposure, capturing both X-rays and electrons.  In contrast to electrons, that leave elongated tracks on the MINIPIX detector, X-rays/gamma-rays give rise to signal confined to single or double pixel on the MINIPIX detector\cite{LOpalka2012,Vavrik2014}. As a result, the track length can be used to separate X-rays and electrons contributions captured in a single acquisition. Figure S8 (B) shows the X-ray image, containing single or double pixel hits only, acquire from Fig. S8 (A), while Fig. S8 (C) shows the corresponding electron image containing track length $\geq$ 5 pixels.\\

\onecolumngrid
\begin{figure}[!h]
\centering
\renewcommand{\thefigure}{S\arabic{figure}}
\includegraphics[width=\textwidth]{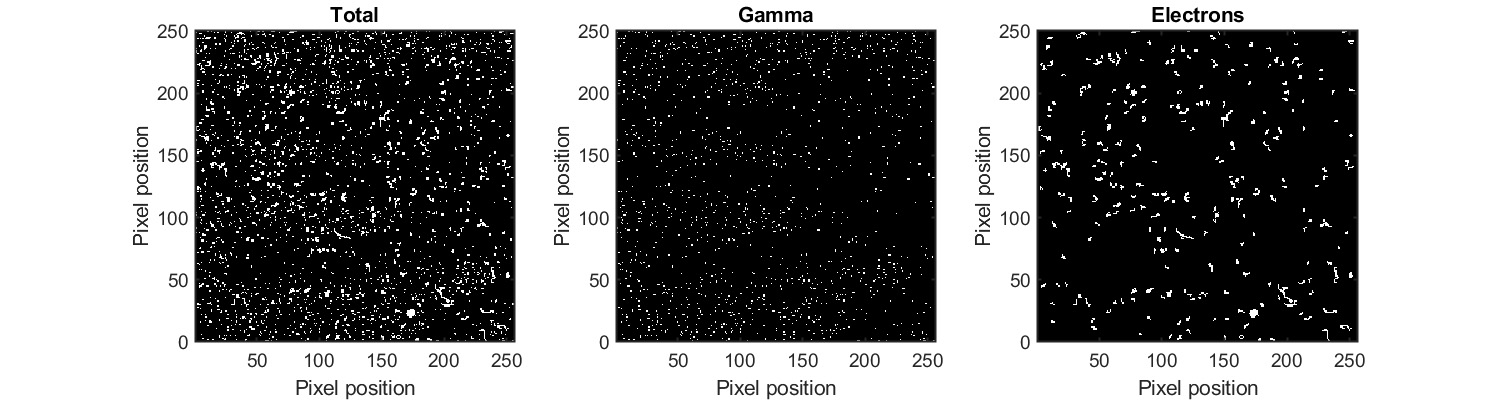}

\caption{ 
 \textbf(A) Total image containing both charged particles and photons, as acquired by MINIPIX in a single frame of  1s exposure at 3$\times$ $10^{15}$ W/cm$^2$. The imaged object is the left segment of the mouse saw shown in Figure 2(C). The red circle highlights a typical elongated track produced by high energy electrons.\textbf(B) Image corresponding to only gamma counts, having track length of 2 pixels or lower. \textbf(C) Image consisting of electrons tracks of track length greater than 5 pixels.}
\end{figure}
\twocolumngrid
\bibliography{sample}